\documentclass[letterpaper,twocolumn,10pt]{article}
\usepackage{usenix}
\usepackage{xcolor}
\usepackage{comment}

\usepackage{booktabs}
\usepackage{tikz}
\usepackage{comment}
\usepackage{amsmath}
\usepackage{caption}
\usepackage{subcaption}
\usepackage{graphicx}
\usepackage{float}
\usepackage{algorithm}
\usepackage{algorithmicx}
\usepackage[noend]{algpseudocode}
\algnewcommand{\LineComment}[1]{\State \(//\) #1}
\usepackage[normalem]{ulem}
\usepackage{bbding}
\usepackage{todonotes}
\usepackage{filecontents}
\usepackage[frozencache,cachedir=.]{minted}
\usepackage{enumitem}
\usepackage[bottom]{footmisc}
\usepackage[]{hyperref}
\usepackage{cleveref}
\usepackage{pifont}
\usepackage{multirow}
\usepackage{xcolor}
\usepackage{outlines}

\definecolor{LightGray}{gray}{0.9}
\definecolor{FadedBanana}{RGB}{255,255,191}
\definecolor{DeepChalk}{RGB}{255,191,191}
\definecolor{FadedFlora}{RGB}{191,255,191}
\definecolor{DeepSnow}{RGB}{191,255,255}
\definecolor{SoapStone}{RGB}{218,218,218}
\definecolor{LightCayenneSixty}{RGB}{239,206,211}
\definecolor{LightCayenne}{RGB}{208,143,145}

\makeatletter
\renewcommand{\paragraph}{%
  \@startsection{paragraph}{4}%
  {\z@}{1.25ex \@plus 1ex \@minus .2ex}{-1em}%
  {\normalfont\normalsize\bfseries}%
}
\makeatother

\newcommand{\sota}{{template-start}}

\newcommand{\sysname}{{T\textsc{idal}}}

\begin{document}

\date{}

\title{\Large \bf Efficient Function-as-a-Service for Large Language Models with \sysname{}}

\author{
{\rm Weihao Cui$^{1,2}$\thanks{Weihao Cui and Ziyi Xu contributed equally to this work.}, Ziyi Xu$^{1}$\footnotemark[1], Han Zhao$^{1}$, Quan Chen$^{1}$\thanks{Quan Chen is the corresponding author}, Zijun Li$^{1}$, Bingsheng He$^{2}$, Minyi Guo$^{1}$}\\
$^{1}$Shanghai Jiao Tong University,$^{2}$National University of Singapore
} 

\maketitle

\begin{abstract}
Large Language Model (LLM) applications have emerged as a prominent use case for Function-as-a-Service (FaaS) due to their high computational demands and sporadic invocation patterns.
However, serving LLM functions within FaaS frameworks faces significant GPU-side cold start.
A fundamental approach involves leveraging a template with function state saved on GPUs to bypass the cold start for new invocations.
Yet, this approach struggles with the high GPU footprint, dynamic initialization behaviors, and lazy GPU kernel loading inherent in LLM functions, primarily due to a lack of insight into the underlying execution details.
In this paper, we introduce \sysname{}, an optimized FaaS framework for LLM applications that achieves fast startups by tracing fine-grained execution paths.
By utilizing the traced execution details, \sysname{} generates adaptive function templates, effectively breaking startup barriers for LLM functions.
Extensive evaluations demonstrate that \sysname{} reduces cold start latency by $1.79\times\text{\textasciitilde}2.11\times$ and improves the $95\%$-ile time-to-first-token by $76.0\%$, surpassing state-of-the-art methods.
\end{abstract}

\section{Introduction}

Function-as-a-Service (FaaS)~\cite{jonas2019cloudprogrammingsimplifiedberkeley} has emerged as the leading serverless paradigm in cloud platforms~\cite{azure_functions,aws_lambda}. By adopting functions as the basic unit of scheduling, FaaS provides distinct benefits for both application developers and cloud providers.
FaaS allows developers to focus on core function logic without caring infrastructure management.
Additionally, its pay-as-you-go model lowers costs, particularly for functions with low invocation frequency~\cite{deloitte_tco_serverless}.
For cloud providers, FaaS enhances resource utilization by enabling more effective resource management, ensuring efficiency and scalability.

Large language model (LLM) applications are emerging as a prominent use case for FaaS.
These applications require both rapid innovation and significant GPU resources but may encounter low invocation frequencies, particularly during early deployment phases.
The high computational demands and unpredictable workloads make FaaS an ideal solution for hosting LLM functions.
Recognizing these benefits, many companies, such as Hugging Face and RunPod~\cite{huggingface_custom_handler,runpod_serverless_application,beam_running_functions_gpu}, now offer FaaS services designed for LLM applications.

We observe that the inherent long cold startup problem in FaaS becomes much worse, exacerbated by the complexity of LLM execution environments.
From memory stack aspect,
executing LLMs requires managing a large footprint across a three-tier memory hierarchy: storage, host memory, and GPU memory.
From application aspect,
different from traditional FaaS applications, costly dynamic model initialization is often required.
For instance, multilingual functions~\cite{chockalingam2024deploy} adapt to individual requests by dynamically attaching language-specific LoRA adapters~\cite{huLoRALowrank} to a shared base LLM~\cite{dubeyLlama3,touvronLlama2,touvronLLaMAOpen}.
The dynamics makes many cache-based optimizations inefficient.
From context preparing aspect,
inference on GPUs necessitates preparing the execution environment, including creating CUDA contexts~\cite{nvidia_cuda_ctx} and other setup operations.

\begin{figure}
    \centering
    \includegraphics[width=0.85\linewidth]{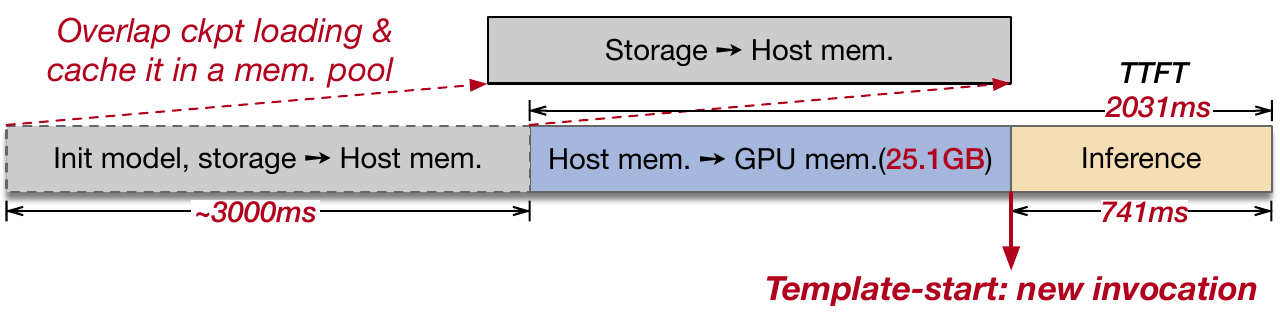}
    \vspace{-2mm}
    \caption{
    Cold-start invocation using Llama2-13B\cite{touvronLlama2} on an Nvidia RTX A6000. The input length is 2k.
    }
    \vspace{-5mm}
    \label{fig:cold-warm}
\end{figure}

\autoref{fig:cold-warm} illustrates the cold start process of an LLM function: initializing the model on the host (loading the checkpoint into host memory), transferring the model from the host to the GPU, and performing the inference on GPU.
The first two steps are the primary contributors to cold start. 
Recent studies~\cite{fuServerlessLLMLowlatency,cheng2023loading} have proposed techniques to mitigate the startup latency associated with host-side initialization.
In the upper half of \autoref{fig:cold-warm}, state-of-the-art frameworks~\cite{fuServerlessLLMLowlatency} overlap storage-to-host model loading with host-to-GPU data transfer and cache the models in a host-side memory pool.
However, even if the storage-to-host transfer (the first step) is optimized to negligible level~\cite{fuServerlessLLMLowlatency}, the time-to-first-token (TTFT) remains $2.74\times$ greater than inference latency, primarily due to host-to-GPU data transfer (the second step).

To entirely eliminate cold start, a fundamental solution is template-start~\cite{liRunDLightweight,duCatalyzerSubmillisecond}.
A template consists of a pre-initialized function state cached in memory.
Template-start allows new invocations to launch directly from the template, bypassing all initialization steps.
To prevent data conflicts, the template must be request-agnostic, and a copy-on-write mechanism is critical for safe template sharing.
As shown in the bottom right of \autoref{fig:cold-warm}, by saving the initialized LLM in GPU memory as a template, we could directly launch a new invocation, entirely excluding the cold-start latency.
Unfortunately, we find that template-start, which blindly reuses an existing function state without understanding the execution details across function invocation, is generally ineffective for LLM functions.

Firstly, template-start struggles to reconcile the need for fast startup with high template density, due to the substantial footprint of LLMs.
E.g., an Nvidia RTX A6000 GPU (48 GB) can accommodate only one template for the function in \autoref{fig:cold-warm}.
With such a low template density, template-start is impractical within a FaaS framework, where multi-tenancy is essential.
Based on the fact that model weights in the template are accessed sequentially (kernel by kernel) during inference~\cite{baiPipeSwitchFast},
we identify the opportunity to overlap host-to-GPU data transfers with inference.
Such overlap potentially reduces the template footprint and increases template density without impacting latency.
However, such overlapping cannot be achieved without knowing the order of weights are accessed in a model during inference. 

Secondly, template-start fails to reduce cold start latency for functions with dynamic initialization.
As previously discussed, LoRA adapters, which are dynamically attached to the base model, are request-specific~\cite{wuDLoRADynamically,chenPunicaMultiTenant,shengSLoRAScalable}.
Although these adapters account for less than $1\%$ of the base model, the initialization of such LLM functions does not meet the criteria for being saved in the template.
Ideally, it is possible to mitigate cold start for dynamic LLM functions, by creating a template with the most reusable initialization.
Such design currently cannot be achieved, lacking a mechanism that identifies and excludes dynamically initialized components.

Thirdly, starting new invocations from a template with model initialized still suffers from cold start overhead.
Our study in \S\ref{sec:GPU-cold-start} also reveals that the inference time during a cold start is still higher than that within a fully warm function state, where the model has been loaded and executed once.
The key factor is that the code segments for kernel execution are lazily loaded onto GPU during the first inference~\cite{nvidia_cuda_module_management}.
Such overhead can be eliminated
if these code segments could be proactively loaded when pre-warming processes for new invocations.
However, without prior knowledge of the specific GPU kernels launched during inference,
such proactive code loading cannot be achieved.

To this end, we introduce \sysname{}, an efficient FaaS framework tailored for LLM applications with fast cold start.
The key insight of \sysname{} is that a detailed understanding of the fine-grained execution paths of an LLM function—spanning initialization and inference—unlocks new opportunities to optimize cold start.
Unfortunately, the fine-grained execution paths required for startup optimizations are implicit within each invocation and cannot be manually exposed by function developers.
\sysname{} integrates a lightweight tracing mechanism to automatically extract these fine-grained execution paths at runtime.
Using adaptive function templates generated from traced execution paths, \sysname{} tackles the cold-start problem for LLM functions by two optimizations: proactive code segment loading and adaptive state forking.
While proactive code segment loading ensures a fully pre-warmed GPU context, adaptive state forking efficiently reuses static components and overlaps model loading with inference.

We implement \sysname{} by extending PyTorch to serve as the runtime for LLM functions.
\sysname{} transparently supports LLM functions wrapped with a wide range of LLM models.
We conducted extensive evaluations of \sysname{} using representative LLMs~\cite{touvronLLaMAOpen,touvronLlama2,dubeyLlama3,opt,gemma,gpt2} of various sizes, both with and without LoRA enabled.
Experimental results show that \sysname{} achieves $1.79\times\text{\textasciitilde}2.11\times$ speedup in cold start latency compared to state-of-the-art solutions.
Furthermore, under real-world workloads, \sysname{} reduces the $95\%$-ile of TTFT by $76.0\%$.
The key contributions are as follows:
\begin{itemize}
[leftmargin=*,topsep=0.2em,itemsep=-0.2em]
    \item We present a detailed analysis of the cold-start overhead of LLM functions on GPUs and highlight the obstacles of directly applying template-start for optimization.
    \item We identify that the primary challenge in optimizing the cold start of LLM functions lies in the unawareness of the fine-grained execution paths underlying invocations.
    \item We build up a lightweight, weight-centric tracing mechanism to automatically expose the fine-grained execution paths without manual effort.
    \item We consolidate optimizations based on the traced fine-grained execution paths to minimize the startup latency of cold LLM function invocations.
\end{itemize}

\section{Background and Motivation}

\subsection{LLMs and Function-as-a-Service}

Function-as-a-Service (FaaS) products~\cite{huggingface_custom_handler,runpod_serverless_application,beam_running_functions_gpu} are popular among cloud vendors to support large language model (LLM) applications. These frameworks provide scalable and cost-efficient solutions, allowing developers to invoke custom LLM functions tailored to specific use cases.

\begin{figure}[b]
    \centering
    \vspace{-4mm}
    \begin{minted}[
    frame=none,
    obeytabs=true,
    framesep=0mm,
    baselinestretch=0.8,
    fontsize=\footnotesize,
    xleftmargin=1.6em,
    breaklines,
    escapeinside=||,
    linenos
]{python}
# ---initialization start---
import torch
# load checkpoints to host memory
llama_weight= torch.load("llama2-13b")
# creat CUDA context
torch.cuda.set_device("cuda:0")
# init model and load it to cuda
llama = Llama()
llama.load_state_dict(llama_weight)
llama = llama.cuda()
# ---initialization end---

def handler(event, context):
    # model inference on GPU
    output = llama(event["input"])
    return {"output": output}
\end{minted}
\vspace{-5mm}
    \caption{An example of an LLM function encapsulating LLaMA 2-13B with two parts: initialization and handler.}
    \label{fig:faas-example}
\end{figure}

Under this context, application developers encapsulate the LLM inference logic within function code.
\autoref{fig:faas-example} illustrates such an example written in Python, adhering to the coding logic employed in several state-of-the-art research works and industry products~\cite{fuServerlessLLMLowlatency,suiPrewarmingNot,runpod_serverless_application,huggingface_custom_handler,beam_running_functions_gpu}.
The code is uploaded to the framework, which automatically invokes the function instance in response to triggered events (e.g., a RESTful http request).
Once the function is triggered, the framework loads the function code, initializes the corresponding LLM models on GPUs, and executes the function handler for inference.
For high request rates, the initialized model is reused to efficiently process subsequent requests .
However, at low request rates, function is started from scratch, requiring time-consuming model initialization for each request
This results in the common issue known as cold startup in serverless computing.

\subsection{Cold Start of LLM Functions}
\label{sec:cold-start}
\begin{figure}
    \centering
    \includegraphics[width=.9\linewidth]{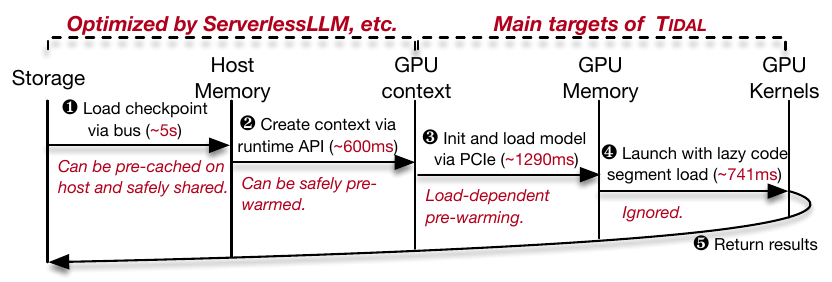}
    \vspace{-3mm}
    \caption{Lifecycle of a cold-start invocation using Llama2-13B and highlighting the optimizing targets of \sysname{}.}
    \label{fig:tidal-targets}
    \vspace{-4mm}
\end{figure}

Researchers have recently focused on alleviating the cold start issues of functions on GPU~\cite{fuServerlessLLMLowlatency,suiPrewarmingNot,liTetrisMemoryefficienta,hongOptimusWarming}.
\autoref{fig:tidal-targets} presents the lifecycle of a cold-start LLM invocation on GPU, utilized to better demonstrate these works and define the target scope of \sysname{}.
In the figure, we omit the CPU-side initialization processes, such as container creation, as these have been extensively explored in CPU-only studies~\cite{duCatalyzerSubmillisecond,fuerstFaasCacheKeeping,yuRainbowCakeMitigatinga,oakesSOCKRapid}.

\paragraph{Warming of model weight on host.}
Stage-1 in \autoref{fig:tidal-targets} primarily involves loading model checkpoints into host memory.
State-of-the-art solutions~\cite{fuServerlessLLMLowlatency} tailored for LLMs utilize a pinned host memory pool to efficiently load and cache model weights from local storage.
With host memory offering significantly greater capacity than GPU memory ($10.7\times$ in our testbed), advanced pre-loading policies~\cite{brookerDemandContainer,liTetrisMemoryefficienta,suiPrewarmingNot} can further reduce the impact of Stage-1 to a negligible level.
Throughout this paper, unless otherwise indicated, we assume that model weights are pre-cached in host memory by existing solutions.

\paragraph{Warming of CUDA context on GPU.}
\label{sec:cuda_context}
Stage-2 in \autoref{fig:tidal-targets} involves the creation of the CUDA context, which initializes essential data structures for GPU-enabled processes.
While this stage is notoriously time-consuming, it is function-agnostic, meaning it can be pre-warmed independently of the function invocations.
As a result, pre-warming the CUDA context with a process pool allows it to be excluded entirely from the cold start of LLM function invocations.

\begin{figure}
    \centering
    \includegraphics[width=0.78\linewidth]{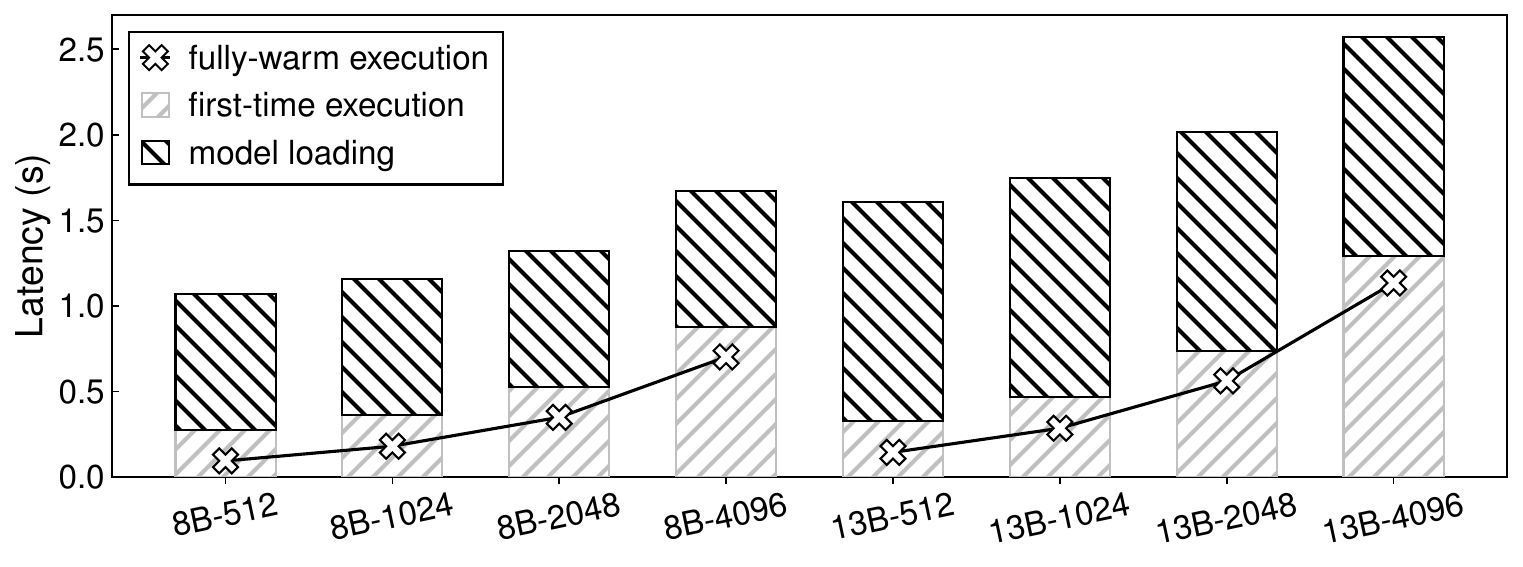}
    \vspace{-2mm}
    \caption{
    Breakdown of GPU cold start and fully-warmed invocation latencies for 2 Llama-family models~\cite{touvronLlama2,dubeyLlama3} with varied inputs.
For instance, ``13B-512'' denotes a Llama with 13 billions parameters evaluated using an input length of 512.
    }
    \label{fig:cold-start-exp}
\end{figure}

\paragraph{Warming of function state on GPU.}
\label{sec:GPU-cold-start}
We collectively term Stages-3\&4 in \autoref{fig:tidal-targets} as the ``GPU cold start'', that remains largely unexplored. 
It involves loading the initialized model into GPU memory and launching GPU kernels for the first-time execution.
To dissect the GPU cold start, we conduct experiments with representative LLMs~\cite{touvronLlama2,dubeyLlama3} on an Nvidia RTX A6000 GPU.
In the experiment, we change the length of input sequences from 512 to 4k for LLMs for simulating various tasks~\cite{patelSplitwiseEfficient,bai2023longbench}.
\autoref{fig:cold-start-exp} illustrates the latency of a cold-start invocation compared to that of an invocation within a fully warmed function state.

Observed from \autoref{fig:cold-start-exp}, Stage-3 requires $2.11\times$ more time than Stage-4 on average.
The long time of Stage-3 is dominated by GPU-side modeling loading through PCIe.
The issue worsens with shorter input sequences and larger model sizes.
Moreover, the latency of Stage-4 exceeds that of a fully warmed invocation by an average of $76.1\%$, equivalent to $179$ ms in absolute terms,
primarily due to the loading of kernel-related code segments into the GPU during the \textit{first-time} kernel execution.
The long GPU cold start emphasizes the critical need for optimization.

Existing works~\cite{suiPrewarmingNot} have tried to optimize Stage-3 by employing pre-warming policy.
The pre-warming method is inherently load-dependent, failing to fundamentally address the cold start caused by model loading over PCIe.
Moreover, to the best of our knowledge, we find no work on eliminating the overhead in Stage-4.
\textit{
Consequently, this paper focuses on systematically optimizing the warming of GPU function state to mitigate the cold start overhead in LLM function invocation.
}

\subsection{Limitations of a Strawman Solution}
\label{sec:strawman_limitations}

\begin{figure}
    \centering
    \includegraphics[width=\linewidth]{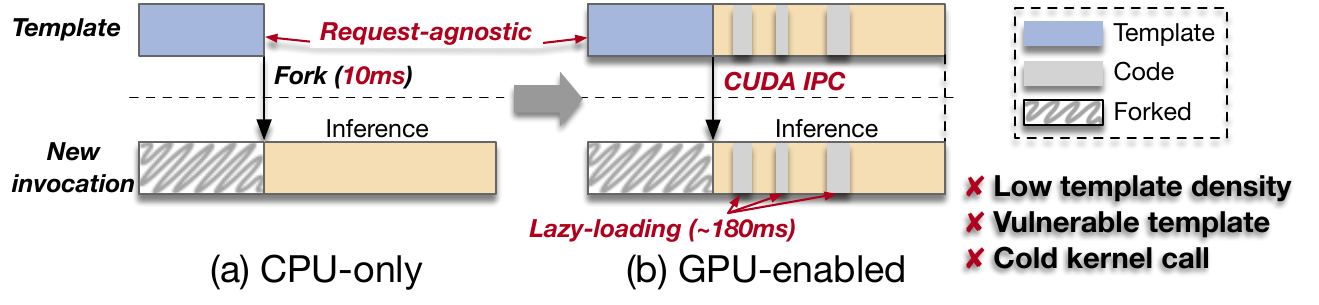}
        \vspace{-5mm}
    \caption{Strawman solution based on CPU-only \sota{}, implemented via CUDA IPC.}
    \vspace{-5mm}
    \label{fig:sfork}
\end{figure}
A strawman solution of achieving warm GPU function states for cold invocations is adopting template-start~\cite{duCatalyzerSubmillisecond,liRunDLightweight}, a method proven effective in the traditional CPU-only FaaS to resolve cold start.
As shown in \autoref{fig:sfork}-(a), this method saves only request-agnostic initializations in a template and leverages the system call to launch new invocations from the template within 10ms.
With data sharing via CUDA inter process communication (CUDA IPC)~\cite{nvidia_cuda_cpp_interprocess}, a similar template for LLM functions can be prepared by first loading the model into GPU.
\autoref{fig:sfork}-(b) illustrates this approach and summarizes its limitations, which stem from GPU or LLM-specific factors and are further elaborated below.

\begin{table}
\footnotesize
    \centering
    \caption{Memory footprint of weights with varied model.}
    \vspace{-2mm}
    \label{tab:model_size}
    \begin{tabular}{c|c|c}
    \hline
    Model & ResNet-101~\cite{heDeepResidual} & Llama Family-8B/13B/34B~\cite{touvronLLaMAOpen,touvronLlama2,dubeyLlama3}\\
    \hline
    \hline
    Size (B) & 170M & 15.7G/24.3G/60G\\
    \hline
    \end{tabular}
    \vspace{-2mm}
\end{table}
\paragraph{Fast startup but low template density.}
The strawman solution requires warming the entire model.
After the warm-up phase, the GPU memory utilized by the template primarily consists of two components: the CUDA context (around 500 MB), and the model weights.
\autoref{tab:model_size} compares the memory footprint of traditional small models with that of popular LLMs.
While the footprint of small models can be even smaller than the requirement for the CUDA context, the situation becomes significantly worse for LLMs due to their substantial memory demands.
As mentioned before, an Nvidia RTX A6000 GPU can support only a single template for Llama2-13B.
Such low template density for fast startup is impractical for a FaaS framework, which is designed to accommodate numerous tenants’ functions.

\begin{figure}
    \centering
    \includegraphics[width=.72\linewidth]{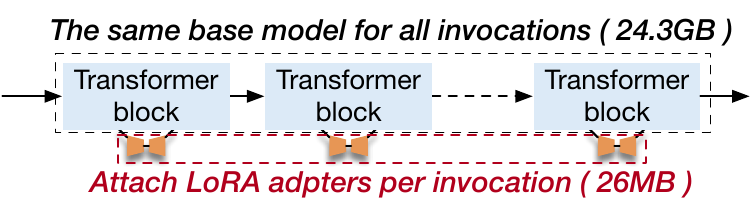}
    \vspace{-2mm}
    \caption{Initializing models with different LoRA adapters for different invocations.}
    \vspace{-5mm}
    \label{fig:lora-llm}
\end{figure}

\paragraph{Vulnerable request-agnostic template.}
Employing the strawman solution requires that the saved function state within the template is request-agnostic.
Failure to do so would result in incorrect outputs and undermine the statelessness principle of FaaS\cite{liServerlessComputing}.
Nonetheless, the model initialization for LLM functions inherently exhibits request-specific variability.
Application developers may utilize parameter-efficient finetuning techniques~\cite{hanParameterEfficientFineTuning}, like LoRA~\cite{huLoRALowrank}, to better meet individual user needs~\cite{wuDLoRADynamically,chenPunicaMultiTenant,shengSLoRAScalable}.
As shown in \autoref{fig:lora-llm}, LoRA~\cite{huLoRALowrank} allows the base model to remain unchanged across invocations while loading user-specific LoRA adapters for each request.
Although adapters are significantly smaller than the base model (<$1\%$), the initialization of such dynamic LLM functions is not request-agnostic and cannot benefit from template-start.
This limitation reduces the general applicability of template-start for resolve cold start.

\paragraph{Non-shareable and lazy-loading code segments.}
Not all GPU data are shareable through CUDA IPC, as only memory explicitly allocated by the process is accessible through this mechanism.
Code segments used for launching kernels during inference fall into this category. The CUDA runtime implicitly and lazily loads these segments during the first invocation of the related kernels.
Thus, the strawman solution still suffers from cold kernel calls (around $180ms$ in \autoref{fig:sfork}).
A straightforward mitigation is to eagerly load all code segments during pre-warming of CUDA context.
Our experiments show that this approach incurs an additional 1.12 GB of GPU memory consumption and increases the pre-warm time for a process from 830 ms to 3050 ms.
This significant overhead makes it impractical to maintain a process pool for LLM invocations.

\subsection{Opportunities \& Challenges}
\begin{figure}
    \centering
    \includegraphics[width=.8\linewidth]{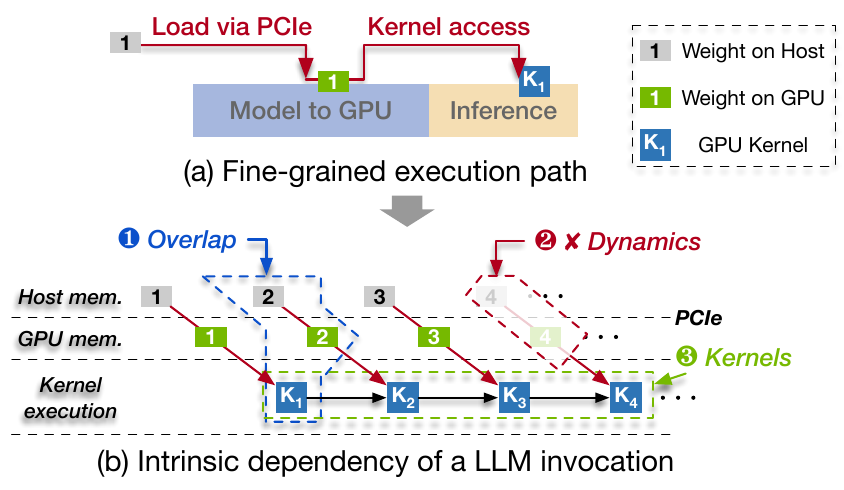}
    \vspace{-2mm}
    \caption{Fine-grained execution path of model weight-1 and intrinsic dependency of an LLM function invocation.}
    \vspace{-4mm}
    \label{fig:exec-path}
\end{figure}

\begin{figure*}
    \centering
    \includegraphics[width=.8\linewidth]{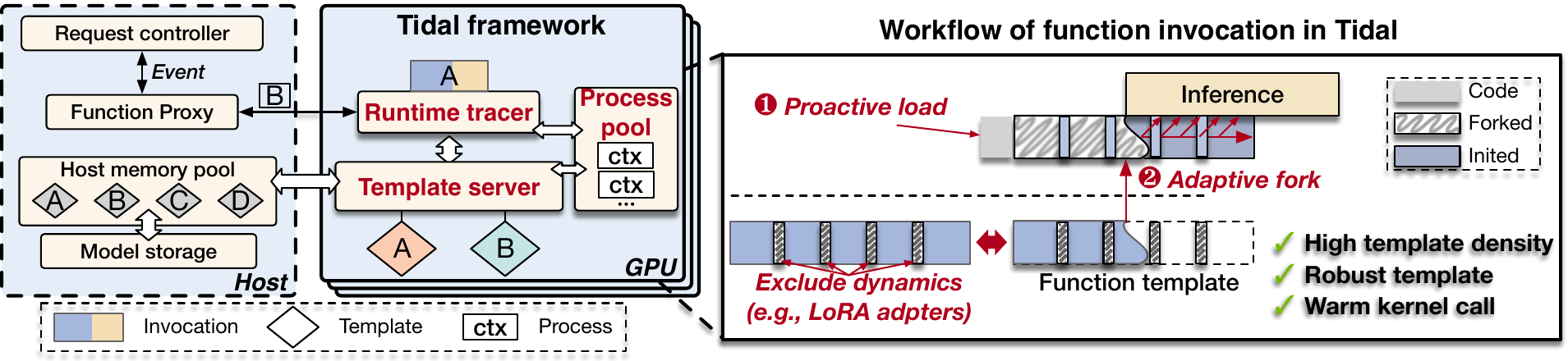}
    \vspace{-1mm}
    \caption{Design overview of \sysname{} and workflow of function invocation in \sysname{}.}
    \vspace{-4mm}
    \label{fig:tidal-overview}
\end{figure*}
\paragraph{Fine-grained execution paths help.}
\autoref{fig:exec-path}-(a) depicts a fine-grained execution path for a specific model weight.
In a framework like PyTorch~\cite{pytorch} for building LLM functions, weights are loaded from host to the GPU as tensors during initialization and subsequently utilized by operators through GPU kernel launches during inference.
Moreover, model inference is also commonly represented as a data flow graph consisting of interconnected operators.
By aligning the execution paths of all weights with the topological order of operators in inference, we derive the intrinsic dependencies of cold-start LLM invocation, as shown in \autoref{fig:exec-path}-(b).

\autoref{fig:exec-path}-(b) reveals three key optimizations to address the limitations. 
1) Kernel execution can overlap with weights loading required for subsequent kernels, removing the need to store the entire model in template.
Since model loading via PCIe and inference latencies are comparable (hundreds of milliseconds), carefully managed overlapping would avoid extra latency while increasing template density.
2) Dynamic elements of model initialization, such as LoRA adapters, can be removed by comparing execution paths across multiple invocations.
As these elements constitute only a small fraction, and the base model's weights are still saved within the template, dynamic LLM functions would benefit from template-start.
3) Kernels specific to an LLM function can be profiled during inference and loaded proactively. Since each LLM function typically uses a small subset of kernels from the kernel libraries, these kernels can be pre-warmed with minimal overhead but excluded from cold start.

\paragraph{Challenges.}
There are several challenges in achieving the above optimizations based on fine-grained execution paths. 
1) Fine-grained execution paths are implicit, making them difficult to be exposed manually by developers;
2) LLM functions have per-invocation dynamic behaviors, requiring runtime extraction of execution paths;
3) General support of these optimizations for various customized LLM functions is hard within a FaaS framework.

\section{Design Overview}
To address these challenges, we introduce \sysname{}, an efficient FaaS framework tailored for LLM applications.
\sysname{} first proposes a lightweight mechanism for transparently tracing fine-grained execution paths.
By leveraging these paths, \sysname{} resolves cold start issues across diverse LLM functions, achieving high template density, robust templates, and pre-warmed kernel calls.

\autoref{fig:tidal-overview} depicts the design of \sysname{}, along with a streamlined function invocation workflow.
\sysname{} comprises three key modules: a \textit{runtime tracer}, a \textit{template server}, and a \textit{process pool}.
Each function invocation runs atop a runtime tracer that traces the fine-grained execution paths with low overhead (\S\ref{sec:runtime_trace}).
It closely interacts with the template server to generate the function templates and launch new invocations.
The template server stores the function templates and retrieves model weights from a pinned host memory pool for each invocation.
The process pool provides pre-warmed processes for new function invocations.

\paragraph{Invocation workflow.}
For each LLM function, a function template is prepared either offline or online, based on fine-grained execution paths.
\sysname{} can adjust the size of model weights cached on the GPU within the template, while the remaining weights are loaded concurrently with inference.
The template is updated by excluding dynamic components at runtime (\S\ref{sec:generate_template}).
Once the template is ready, \sysname{} accepts user requests.
Using the template, it pre-warms processes by initializing CUDA contexts and loading kernel code segments proactively (\S\ref{sec:proactive_load}).
Upon invocation, \sysname{} adaptively forks the saved LLM from the function template as follows:
CPU operations for static model architectures are skipped,
dynamic components are initialized as required,
and weights not cached on the GPU are loaded asynchronously.
When overlapping model loading with inference, \sysname{} provides strict correctness guarantees.
(\S\ref{sec:adapative_fork}).
\begin{figure}
    \centering
    \begin{minted}[
    frame=none,
    obeytabs=true,
    framesep=0mm,
    baselinestretch=0.8,
    fontsize=\footnotesize,
    xleftmargin=1.8em,
    breaklines,
    escapeinside=||,
    linenos
]{python}
import tidal

@tidal.init(static=False)
def initializer(event, context):
    ...
    llama_weight = tidal.load(event["llama2-13b"])
    lora_weight = tidal.load(event["lora"]) # dynamic
    all_weights = llama_weight + lora_weight
    llama_lora = LlamaLoRA()
    llama_lora.load_state_dict(all_weights)
    llama_lora = llama_lora.cuda()
    return llama_lora

def handler(event, context):
    llama_lora = initializer(event, context)
    output = llama_lora(event["input"])
    return {"output": output}
\end{minted}
\vspace{-5mm}
    \caption{Programming interface of \sysname{}.}
\vspace{-4mm}
    \label{fig:programming-interface}
\end{figure}
\paragraph{Programming interface.}
\sysname{} adheres closely to the programming model of modern FaaS frameworks, as shown in \autoref{fig:faas-example}.
To express an LLM function in a traceable manner, \sysname{} imposes minimal requirements on application developers.
\autoref{fig:programming-interface} illustrates an example of defining a function using Llama2-13b with LoRA enabled in \sysname{}.
The most noticeable difference is that model developers need to wrap the function initialization within a method decorated with \texttt{tidal.init} and explicitly invoke it within the function handler,
This requirement arises because an LLM model with dynamics must be initialized for each invocation.
The \texttt{tidal.init} decorator accepts an optional argument, \texttt{static}, to indicate whether the initialization is static or dynamic.
Without annotation, \sysname{} defaults to treating it as dynamic.

\section{Template Based on Fine-grained Tracing}

In this section, we first introduce \sysname{}’s lightweight tracing mechanism, which transparently extracts fine-grained execution paths.
Additionally, we show details of generating function templates for startup optimizations.

\subsection{Tracing Fine-grained Execution Paths}
\label{sec:runtime_trace}

\paragraph{Traceability of LLM functions.}
Fundamentally, LLMs are large-scale deep learning models targeting natural language processing tasks.
Therefore, LLMs follows the generic deep learning paradigms, where model inference is abstracted as a data flow graph composed of tensors and operators~\cite{maRammerEnabling}.
While prior works~\cite{pytorch_dynamo_overview} have leveraged this property to trace inference, \sysname{} further extends it to model initialization.
This extension is driven by the observation that, once loaded into host memory, a weight is represented as a tensor, and subsequent operations performed on it--such as transferring weights from CPU to GPU--are all implemented as deep learning operators.
Fortunately, these traceable operations correspond to the fine-grained execution paths that \sysname{} needs to optimize the startup process.
\sysname{} leaves other non-traceable CPU operations during initialization to be executed as normal.

\begin{figure}
    \centering
    \includegraphics[width=.9\linewidth]{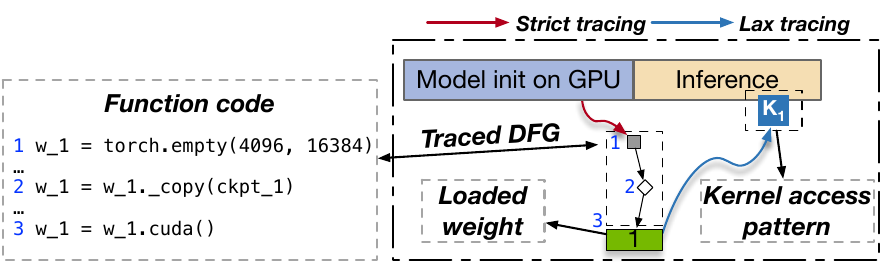}
    \vspace{-1mm}
    \caption{Weight-centric two-phase tracing for weight-1.}
\vspace{-4mm}
    \label{fig:runtime-trace}
\end{figure}

\paragraph{Weight-centric two-phase tracing.}
Blindly tracing function invocation involves going through all tensors and operators, even those used for temporary storage or reshaping during inference.
This leads to considerable runtime overhead.
Our preliminary results show that the overhead can be several times greater than that of a standard cold-start invocation.
Since \sysname{} focuses on optimizing cold starts in LLM function invocations, and model weight loading dominates the initialization, there is no need to treat initialization and inference with the same tracing mechanism.

Therefore, \sysname{} adopts a weight-centric two-phase tracing approach, as illustrated in \autoref{fig:runtime-trace}.
This method focuses on execution paths related to weights.
During model initialization, \sysname{} employs strict tracing to construct data flow graphs (DFGs) for generating weight tensors. Each weight tensor’s DFG specifies the shape used for initialization, the checkpoint for loading weight data, and other relevant details.
Using the DFG, \sysname{} identifies weights that are dynamically initialized. For example, in an LLM function with LoRA enabled, the adapters are flagged as dynamic because they are sourced from different checkpoints, although their shape remains the same.
During inference, \sysname{} employs lax tracing, capturing only the access patterns of weights, including their order and associated GPU kernels.
The \texttt{tidal.init} decorator in \autoref{fig:programming-interface} enables \sysname{} to differentiate between the two phases, activating the appropriate tracing mechanism.

\subsection{Generating Adaptive Function Template}

\label{sec:generate_template}
\begin{figure}
    \centering
    \includegraphics[width=.62\linewidth]{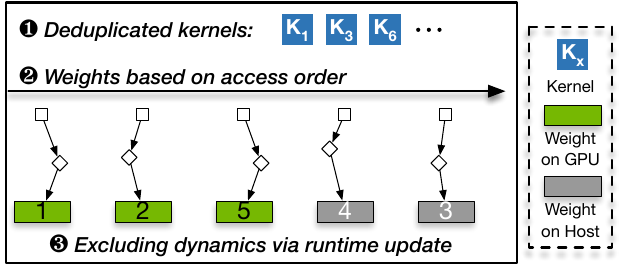}
    \vspace{-1mm}
    \caption{Adaptive function template composed of three key components. The numerical values indicates the sequence in which the weights are initialized.}
    \vspace{-4mm}
    \label{fig:function-template}
\end{figure}

With the traced fine-grained execution paths, \sysname{} generates the function templates for LLM functions managed by the template server.
The function template generated is adaptive in several key aspects:
first, it perceives the necessary kernels for proactive code segment loading with minimal overhead;
second, it adapts its saved state to accommodate dynamic LLM initialization;
third, its GPU memory consumption could be adjusted with loading efficiency guarantee.
Such adaptiveness enables \sysname{} to address cold starts for various LLM functions, regardless of whether models are dynamically initialized, or what workloads are processed (e.g., model sizes, request rates, input lengths, etc.).
As shown in \autoref{fig:function-template}, the template for a specific LLM function is generated as follows.

Firstly, information about GPU kernels required for proactive loading of code segments is stored in the template.
Since LLM models often consist of multiple identical transformer blocks, \sysname{} scans all traced operators and filters them by removing duplicates.
GPU kernels associated with these operators are then identified as candidates for proactive loading.

Secondly, model weights are stored in the template with a re-ordered memory layout. \sysname{} reorganizes the weights based on the traced access order. Without this reorganization, our evaluation in \S\ref{sec:eval:optimization} reveals that overlapping efficiency can be easily compromised by misordered weight initialization, highlighting the necessity of \sysname{}’s tracing mechanism. By leveraging the traced access order, \sysname{} also retains a subset of model weights on the GPU while preserving only the memory layouts of others, thereby optimizing the template size. Weights stored as memory layouts in the template are efficiently loaded into the GPU during inference.

Thirdly, data flow graphs for generating each model’s weights are stored in the template to facilitate the exclusion of dynamic components at runtime. Since \sysname{} cannot identify all dynamic components of an LLM model in a single tracing pass, its low-overhead tracing mechanism enables the incremental exclusion of these components during runtime.

\section{Optimizations of Function Startup}
In this section, we introduce the key startup optimizations in \sysname{} based on insights from fine-grained tracing: proactive code segment loading, and adaptive state forking.

\subsection{Proactive Code Segment Loading}
\label{sec:proactive_load}
Avoiding cold kernel call startup requires knowledge of the kernels to be launched during inference.
As discussed in \S\ref{sec:strawman_limitations}, template-start incurs a 180-millisecond cold start due to the absence of this information.
\sysname{} addresses it by leveraging traced kernels stored in its template to proactively load the relevant code segments during process pre-warming.
As a result, new invocations do not suffer cold kernel calls.

However, this process requires launching the corresponding kernels on the GPU to trigger proactive loading, which also occupies computational resources.
To minimize interference with other ongoing invocations, \sysname{} reduces the input dimensions of the triggering kernels.
Additionally, as GPU kernels are already deduplicated during template generation, the kernel code segments are efficiently loaded during pre-warming with minimal computational and memory overhead, as evaluated in \S\ref{sec:eval:optimization}.

\paragraph{Loading policy.}
Another issue arises from the multi-tenancy of FaaS, where a single GPU instance serves multiple LLM functions.
Proactively loading code segments for all deployed functions within a single process may lead to excessive GPU memory usage and increased triggering latency.
To address this, \sysname{} employs a proactive loading policy that aligns with the set of LLM functions currently cached in the host memory of the GPU instance.
During the process pool’s pre-warming phase, \sysname{} triggers the loading of all the deduplicated kernels corresponding to the templates for these cached functions.
This policy ensures that the code segments of frequently accessed kernels are readily available.

\subsection{Adaptive State Forking}
\label{sec:adapative_fork}
With the process pre-warmed with loaded code segment, \sysname{} employs adaptive state forking to efficiently initialize function states from the template for new invocations.

\paragraph{Adaptive template state reusing.}
When an LLM is dynamically initialized based on user requests, the function initialization cannot be saved into the template for fast startup in template-start. As shown in \autoref{fig:exec-path}, \sysname{} maximizes reuse of static elements from the existing state within the template by leveraging its traced fine-grained execution paths. Since dynamic elements typically account for only a small portion of the entire model (barely $1\%$), \sysname{} achieves fast startup by reusing the majority of the initialized LLM.

Specifically, \sysname{} supports both dynamic initialization and template-based startup through its runtime tracing.
Instead of directly reusing function state from template, \sysname{} executes function initialization atop of runtime tracer.
For model weight identified as static initialization, \sysname{} reuses it from the template server.
For model weight identified as dynamic initialization, \sysname{} initializes it within user code.

\begin{figure}
    \centering
    \includegraphics[width=\linewidth]{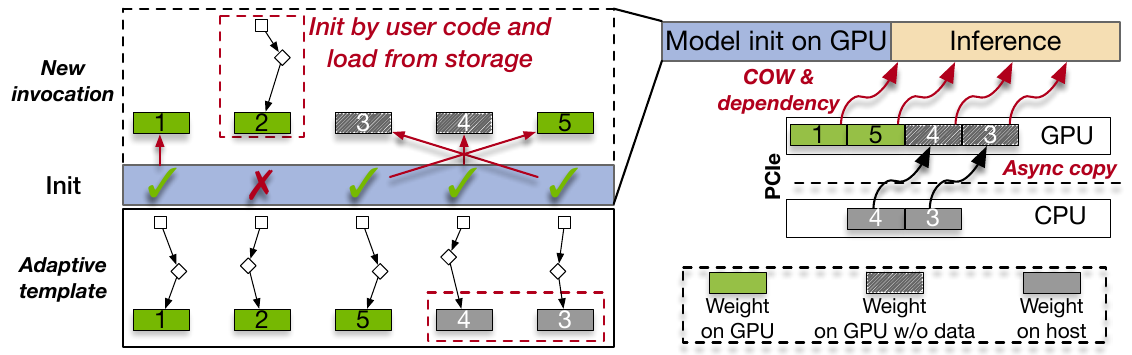}
\vspace{-1mm}
    \caption{Adaptively forking an LLM function invocation and performing inference with overlapped data loading.
    The function is dynamically initialized with LoRA enabled.}
    \label{fig:adaptive-fork}
\vspace{-4mm}
\end{figure}

The left part of \autoref{fig:adaptive-fork} presents an example of how \sysname{} forks a new invocation from its function template.
During initialization, \sysname{} traces the operators involved in GPU weight initialization, generating a data flow graph for comparison with the pre-saved graphs in the function template.
For weights (e.g., weight-1, weight-5, weight-4, and weight-3) whose data flow graphs match the template, \sysname{} skips operator execution and directly initializes the tensors by forking GPU memory pointers from the template.
In contrast, weight-2, loaded from a different adapter, has a data flow graph inconsistent with the template.
As a result, \sysname{} replays the operators for weight-2 to dynamically initialize it.

Dynamic elements, such as LoRA adapters, are handled within user code and can be loaded either from host memory or storage.
A function with LoRA enabled may be equipped with thousands of request-specific adapters~\cite{shengSLoRAScalable,chenPunicaMultiTenant,wuDLoRADynamically}.
Due to this specificity, the caching policies of FaaS frameworks often fail to effectively cache checkpoints for these adapters.
To ensure a fair comparison in \S\ref{sec:eval}, \sysname{} currently loads these dynamically initialized adapters directly from storage.

\paragraph{Efficient overlapping with correctness ensuring.}
On the left side of \autoref{fig:adaptive-fork}, when weights such as weight-3 and weight-4 are forked from the template, only their GPU memory addresses are allocated initially.
At this point, their actual data temporarily resides in the host memory pool and \sysname{}’s template server is loading these weights into the GPU asynchronously in the background.
In this case, \sysname{} is capable of overlapping model loading with inference, thus reducing TTFT of an cold-start LLM invocation to the latency of either loading or inference, whichever is longer.

While template server enables overlapping, two key issues remain: efficiency and correctness.
The misalignment between the order of weight initialization and their access order reduces overlapping efficiency when weights are loaded arbitrarily.
\sysname{} leverages the traced access order to ensure weights are loaded in the correct sequence.
As shown in the right of \autoref{fig:adaptive-fork}, although weight-3 is initialized before weight-4, the template server loads weight-4 ahead of weight-3 because weight-4 is consumed by a kernel first during inference, according to the traced access pattern.

For correctness, \sysname{} addresses two key aspects.
Firstly, dependencies between weights and kernels must be preserved, as data is transferred to the GPU using asynchronous copy operations. To ensure the required data is available on the GPU before kernel execution, \sysname{} injects synchronization events based on the traced execution paths.
Secondly, \sysname{} employs a copy-on-write mechanism to prevent modifications to weights forked for a new invocation. The runtime tracing in \sysname{} inspects the read-write properties of operators before launching the underlying kernels. If an operator attempts to write to a forked weight, \sysname{} copies the weight into a new tensor to maintain its read-only property during inference.

\paragraph{Adapting template size for less cold start.}
As mentioned earlier, \sysname{} reduces the TTFT of a cold-start LLM invocation to the greater of the loading or inference latency.
Ideally, this latency could be reduced to match inference, as inference is the only step that cannot be performed in advance.
\sysname{} could adapt the template size on GPU--the model weights prefetched on the GPU within the template--based on traced model access order to achieve this.
As highlighted in prior works~\cite{baiPipeSwitchFast}, this is fundamentally a trade-off between computation and data transfer.
To improve overlapping, \sysname{} dynamically adjusts the template size based on the following principles.
First, \sysname{} need to analyze the function workloads to determine the average input length and batch size of LLM function requests.
Using the analyzed data, it profiles the warm execution to obtain an average TTFT for the current LLM function.
\sysname{} then adapts the template size according to \autoref{eq:template-size},
\begin{equation}
M_{prefetch} = \max(M_{model} - T_{TTFT} \times B_{PCIe}, 0)
\label{eq:template-size}
\end{equation}
where, $M_{prefetch}$ denotes the template size, $M_{model}$ the entire LLM weight footprint, $T_{TTFT}$ the analyzed average TTFT, and $B_{PCIe}$ the bandwidth of PCIe.
Notably, $M_{prefetch}$ represents the maximum required template size for optimal performance.
\sysname{} dynamically adapts this value using its runtime tracing to balance cold start reduction with high template density.

\paragraph{Keep-alive of dynamic function.}
\label{sec:keep_alive}
FaaS frameworks typically keep a launched function instance alive for a predefined interval to handle subsequent requests and avoid cold starts.
However, for dynamic functions that initialize different models per request, the initialized models cannot be reused for subsequent requests.
With adaptive fork, \sysname{} retains static model weights on the GPU during initialization while only re-initializing dynamic components.
This allows dynamic functions in \sysname{} to benefit from the performance enhancements of keep-alive.
To enable adaptive support for keep-alive across all LLM functions, \sysname{} requires application developers to specify whether a function is dynamically or statically initialized, as mentioned in \autoref{fig:programming-interface}.
During the keep-alive interval, this information allows \sysname{} to skip initialization for static function invocations and leverage adaptive state forking for dynamic ones.
Without this specification, \sysname{} assumes all functions are dynamic, leading to inefficient keep-alive management for static functions.

\section{Implementation}
\sysname{} is implemented in Python and C++ with approximately $5,800$ lines of code. Its function runtime and programming interface are built by extending PyTorch, utilizing $3,900$ lines of C++ and $1,050$ lines of Python.
Although \sysname{} primarily addresses the cold start challenge for individual LLM functions by tracing their fine-grained execution paths, we have also developed a FaaS scheduler prototype to evaluate its performance with real-world workloads.
This scheduler, consisting of 840 lines of Python, supports essential features required in a FaaS cluster, such as keep-alive for functions with high invocation rates and early-reject mechanisms for timeout requests, etc.

\paragraph{Dispatch-based runtime tracing.}
\sysname{}’s tracing mechanism leverages the dispatch mechanism of PyTorch~\cite{pytorch_extending_modes}, which enables the injection of tracing code through its Python frontend before the execution of operators.
Although the weight-centric two-phase tracing is designed to capture only the fine-grained execution paths needed for startup optimization, a naive implementation in Python end introduces substantial overhead due to frequent context switches between Python and C++.
During inference, this overhead can reach as high as $100\%$.
To mitigate this issue, \sysname{} registers its runtime tracer as a custom backend in PyTorch, operating entirely within C++ context.
As a result, \sysname{} eliminates the overhead caused by thousands of context switches, significantly improving the efficiency of tracing and execution.

\paragraph{Tailored memory pool in template server.}
\sysname{}’s template server is designed to efficiently transfer model weights from host memory to the GPU. When overlapping host-to-GPU data transfers with kernel execution, a single LLM may require the transfer of numerous weight tensors.
Transferring these tensors individually, as stored in the template, could saturate the command queue for memory copy operations~\cite{hanMicrosecondscalePreemption}, leading to significant overhead during function initialization.
To address this, \sysname{}’s template server automatically merges weight tensors into fewer tensors when their total number exceeds a threshold.

\section{Evaluation}
\label{sec:eval}

\subsection{Experimental Setup}

\paragraph{Testbed.}
We evaluate \sysname{} with two setups.
The first testbed has four servers, each equipped with an AMD EPYC 7R32 CPU and two Nvidia RTX A6000 (48GB) GPUs.
Each server has 512GB of host memory and communicates with the GPUs via PCIe 4.0, providing a bandwidth of 32GB/s.
The second testbed is a server with an Intel Xeon Platinum 83698 CPU with 8 Nvidia Amper 100 (80GB) GPUs.
This server has 1TB of host memory, and GPU communication occurs through PCIe 3.0, offering a bandwidth of 16GB/s.
Both servers run CUDA-12.5 and PyTorch-2.4.
We use the first testbed for most evaluations including startup of single LLM function and application of \sysname{} within a FaaS cluster.
The second sever is used for evaluating the scalability of \sysname{} for distributed LLM functions.

\paragraph{Benchmarks.}
We evaluate \sysname{} using models from four representative LLM families: GPT-2~\cite{gpt2}, OPT~\cite{opt}, Gemma~\cite{gemma}, and Llama~\cite{touvronLLaMAOpen,touvronLlama2,dubeyLlama3}, covering parameter sizes ranging from 1.5 billions to 70 billions.
For single-GPU LLM function evaluations, we include GPT-2-1.5B, OPT6.7B, Gemma-9B, Llama3-8B, and Llama2-13B.
For evaluation in a distributed inference environment, we utilize Llama2-13B, Llama2-34B, and Llama3-70B.
LoRA adapters are attached to evaluated LLMs for constructing dynamic LLM functions, following the previous works~\cite{wuDLoRADynamically,chenPunicaMultiTenant}.
Across all evaluated LLM functions, we report their time-to-first-token (TTFT), which reflects the startup latency corresponding to Stages 3\&4 in \autoref{fig:tidal-targets}, the primary optimization target of \sysname{}.

\subsection{Startup of LLM Functions}
\subsubsection{TTFT across Various LLM Functions}

\begin{figure}
    \centering
    \includegraphics[width=\linewidth]{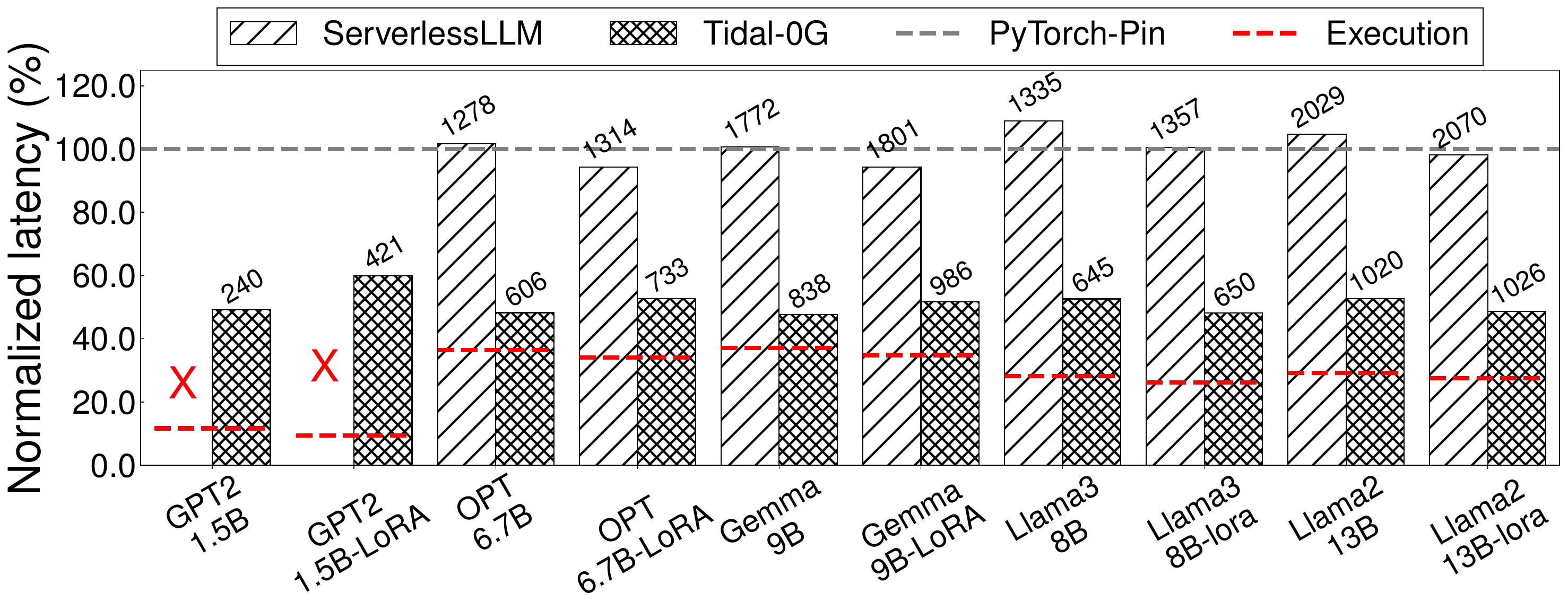}
    \vspace{-7mm}
    \caption{TTFT of LLM functions across different LLMs with input length fixed at 2048 and batch size fixed at 1, including variants with LoRA enabled.}
    \vspace{-4mm}
    \label{fig:eval:ttft-all}
\end{figure}

To the best of our knowledge, \sysname{} is the first approach to introduce template-start in GPU-related applications.
No prior work has demonstrated the capability to launch an invocation directly from a template residing on a GPU.
Consequently, we evaluate \sysname{} against the following three baselines for cold-start LLM function invocation. 
    \texttt{PyTorch-pin} assumes that the model has been pre-initialized in host pinned memory, representing the upper bound of latency for a cold-start LLM invocation.
    \texttt{ServerlessLLM}, state-of-the-art FaaS solution for LLM, caches model weights in a separate host-side pinned memory pool for launching a LLM invocation.
    \texttt{Execution} assumes the model has already been loaded into GPU memory and executed once, representing the lower bound of latency for a cold-start LLM invocation.
In this experiment, \sysname{} prefetches none of weights within the template generated for all evaluated LLM functions for fair comparison (\texttt{Tidal-0G}).
Each LLM is tested in two versions: the original and a LoRA-enabled dynamic variant. The input length is fixed at 2048, and the batch size is set to 1.

\autoref{fig:eval:ttft-all} shows the TTFT of all evaluated LLM functions.
All latencies are normalized to \texttt{Pytorch-pin}.
Compared to \texttt{PyTorch-pin} and \texttt{ServerlessLLM}, \texttt{Tidal-0G} achieves $1.96\times$, $2.00\times$ speedup of TTFT on average, respectively.
While none of the systems occupy GPU memory before function invocation, \sysname{} minimizes startup latency by skipping the initialization of static model elements, overlapping model loading and inference, and proactively loading code segments using fine-grained execution paths traced during runtime.
In contrast, \texttt{PyTorch-pin} and \texttt{ServerlessLLM} require the model to be fully initialized and loaded into GPU memory before inference can begin.
Additionally, their model inference processes are impacted by cold kernel calls, further contributing to latency.
When LoRA is disabled, \sysname{} delivers an average TTFT speedup of $2.07\times$ compared to \texttt{ServerlessLLM}. With LoRA enabled, the speedup decreases to $2.0\times$ on average.
The performance slowdown is expected, as \sysname{}’s runtime tracer cannot trace or save the dynamic initialization of LoRA adapters into the function template.
The initialization of dynamic elements is fully managed by LLM functions themselves without acceleration.
Despite this, by reusing the majority of the initialized model ($99\%$), \sysname{} still achieves significant reductions for dynamic functions.

Compared to \texttt{Execution}, \texttt{Tidal-0G} remains $22\%\text{\textasciitilde}84\%$ slower across all LLM functions.
This slowdown can be attributed to two main factors.
First, all model weights in \texttt{Tidal-0G} are loaded from host memory.
By prefetching a portion of the model weights into \sysname{}’s template (template size), \sysname{} can further approach the lower bound.
The impact of template size will be discussed later.
Second, \sysname{} is unable to trace pure CPU-based operations during initialization, which fall outside its acceleration capabilities.
This limitation is noticeable in the case of GPT2-1.5B, whose initialization involves numerous CPU operations.

Notably, \texttt{ServerlessLLM} fails to execute LLM functions wrapped with GPT2-1.5B.
This limitation stems from the fact that \texttt{ServerlessLLM} is not a native FaaS framework tailored for LLM functions.
\texttt{ServerlessLLM} requires manual adaptation of the LLM model initialization for efficient host-to-GPU data transfer.
In contrast, \sysname{} leverages its runtime tracer to transparently enable its optimizations without the need for manual intervention.
\begin{figure}
    \centering
    \includegraphics[width=.85\linewidth]{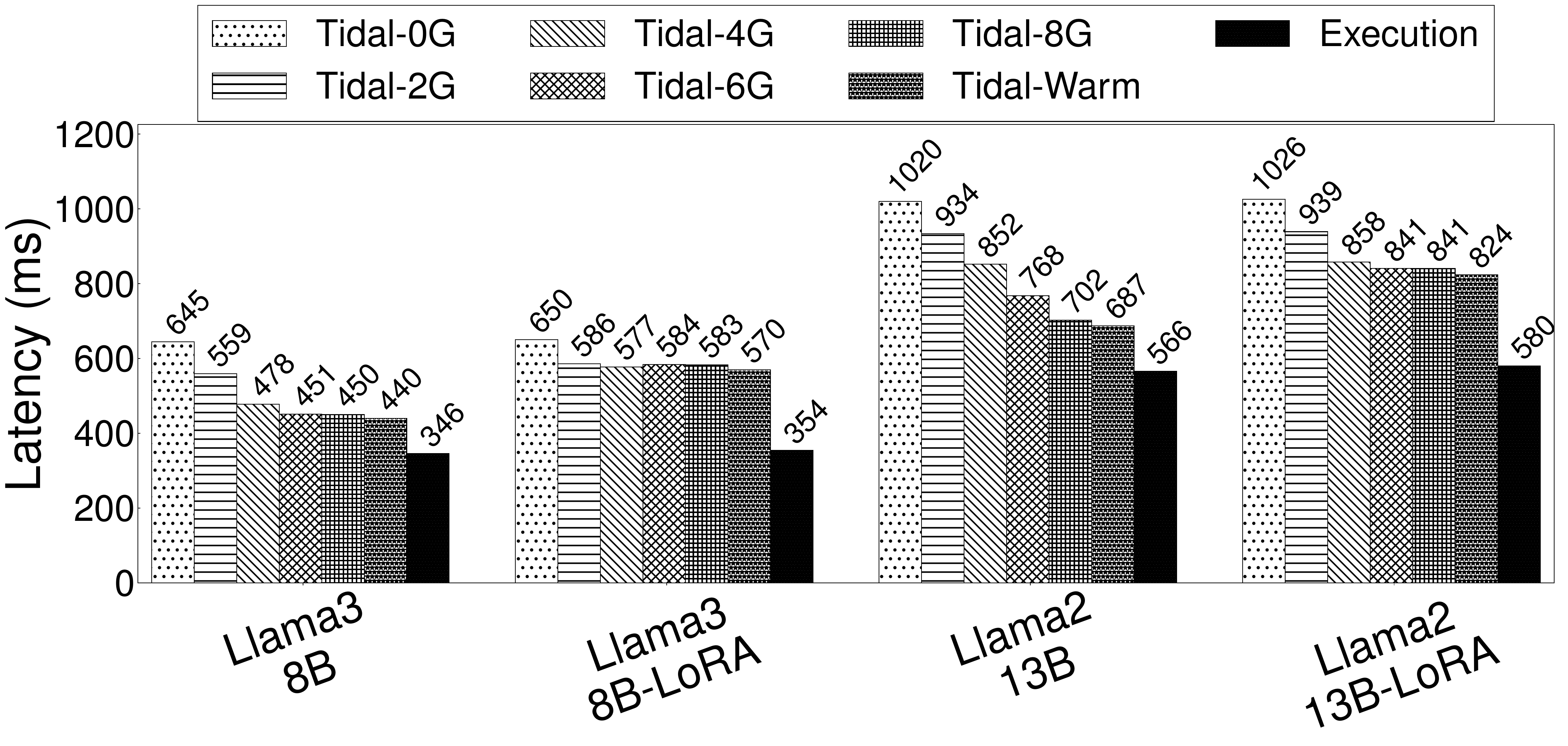}
    \vspace{-4mm}
    \caption{TTFT of Llama-family functions with varied template sizes, including LoRA-enabled variants. Input length and batch size are fixed at 2048 and 1, respectively.
    }
    \vspace{-4mm}
    \label{fig:ttft-template-size}
\end{figure}
\subsubsection{Ablation Study}
\label{sec:eval:abation}
To further analyze \sysname{}’s effectiveness in reducing startup latency for LLM functions, we conduct ablation studies.
The following experiments focus on Llama-family models, as they are among the most widely used pretrained LLMs for customization.
Experimental results with other LLMs demonstrate similar trends.
We omit the results of \texttt{PyTorch-pin} and \texttt{ServerlessLLM} from most experiments, as they are significantly slower than all configurations of \sysname{}.

\paragraph{TTFT with varied template sizes.}

In this experiment, we evaluate \sysname{}’s effectiveness by varying the template size on the GPU. \autoref{fig:ttft-template-size} illustrates the TTFT of the evaluated LLM functions as the template size varies from $0G$ to the entire model size.
Specifically, \texttt{Tidal-Warm} refers to the configuration where the template size equals the entire model size.
Compared to \texttt{Tidal-0G}, \texttt{Tidal-Warm} achieves a $14\%\text{\textasciitilde}48\%$ TTFT speedup across all evaluated functions.
Generally, \sysname{}’s performance improves with larger template sizes, up to the point where model loading fully overlaps with model inference.
From the figure, we also observe that the template size required for dynamic LLM functions with LoRA enabled to achieve the best TTFT is smaller than that of static functions.
This is because the initialization of dynamic functions takes longer than their static counterparts, allowing \sysname{} to use this time to overlap more model loading.

\begin{figure}
    \centering
    \includegraphics[width=.95\linewidth]{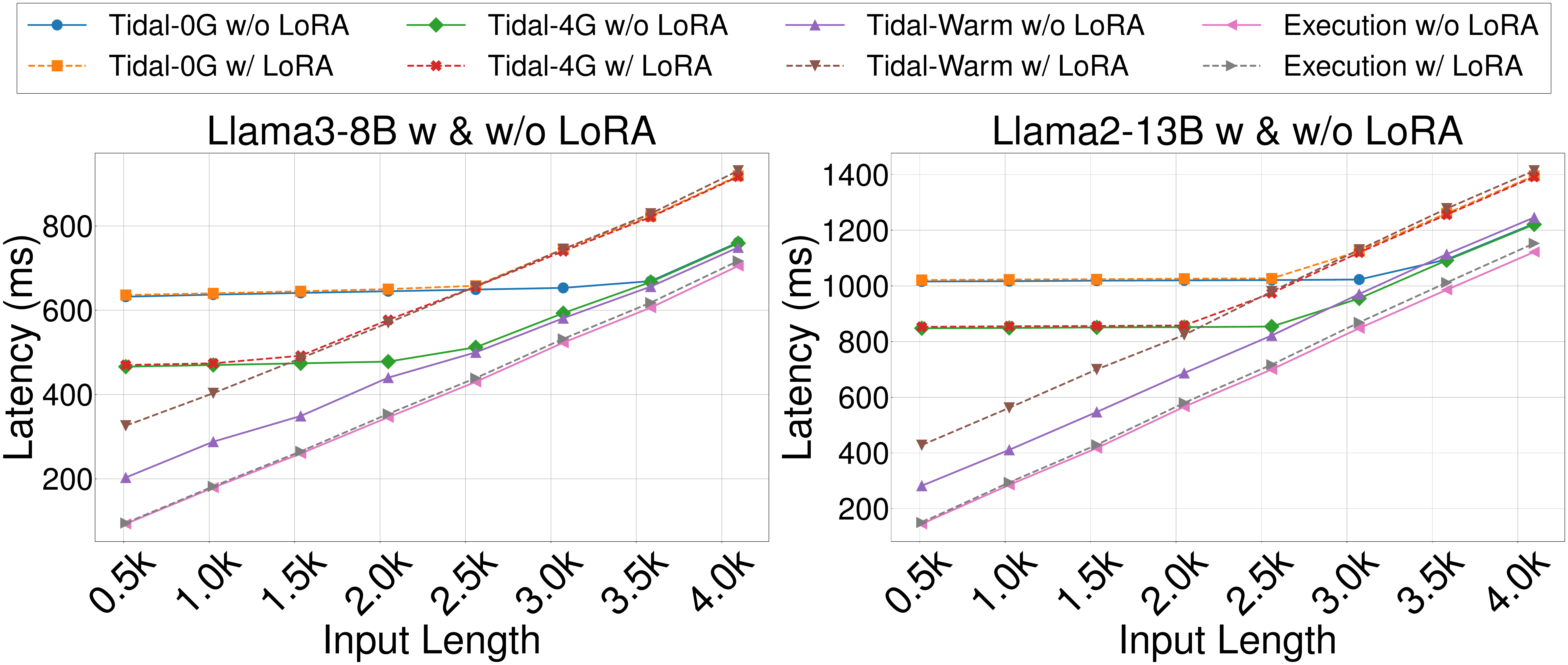}
        \vspace{-2mm}
    \caption{TTFT of Llama-family functions with varied input lengths and template sizes (0G, 4G, and entire model), including LoRA-enabled variants. The batch size is set to 1.
        \vspace{-1mm}
    }
    \label{fig:ttft-len}
\end{figure}
\begin{figure}
    \centering
        \includegraphics[width=.95\linewidth]{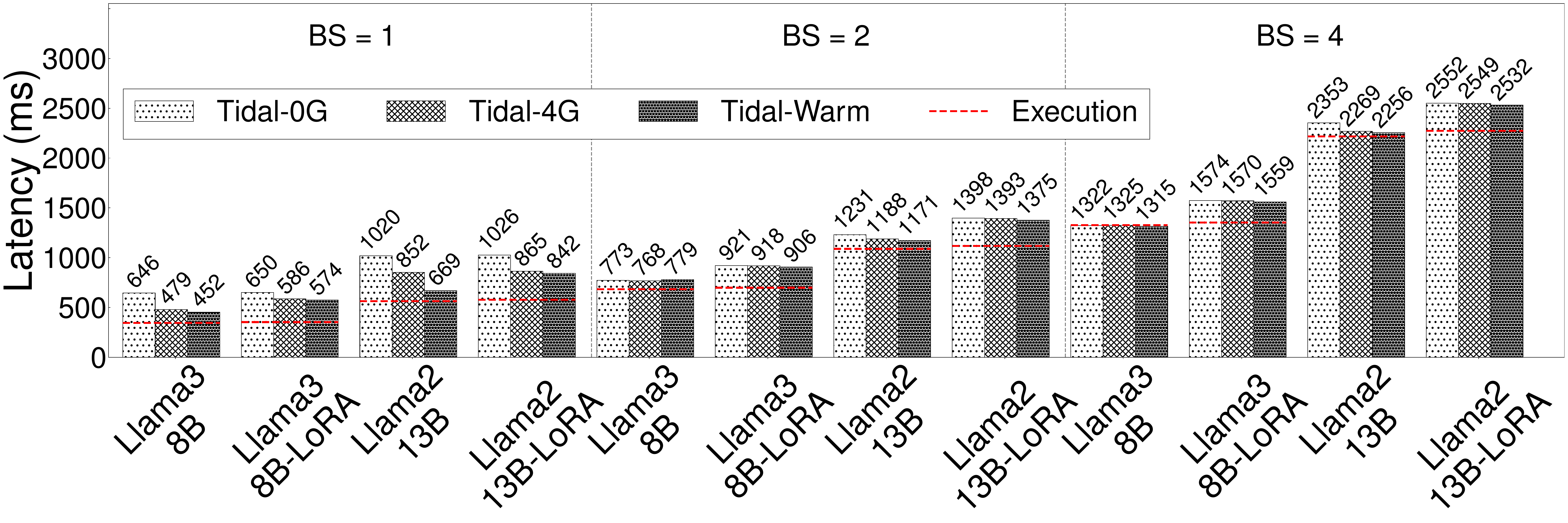}
        \vspace{-3mm}
        \caption{TTFT of Llama-family functions with varied batch sizes and template sizes (0G, 4G, and entire model), including LoRA-enabled variants. The input length is set to 2048.}
        \vspace{-4mm}
        \label{fig:ttft-bs}
\end{figure}
\paragraph{TTFT with varied input lengths and batch sizes.}
In this experiment, we explore the impact of function workloads on \sysname{}’s performance. To evaluate this, we create different workloads by varying function input lengths and batch sizes. \autoref{fig:ttft-len} illustrates the TTFT of the evaluated functions with varying input lengths, while \autoref{fig:ttft-bs} presents the TTFT of the evaluated functions with increased batch sizes. Both figures include three variants of \sysname{}, each configured with a different template size.
From these figures, we observe a turning point for \texttt{Tidal-0G} and \texttt{Tidal-4G}, where their TTFT converges with that of \texttt{Tidal-Warm} when the input length or batch size exceeds a certain threshold.
This behavior occurs because higher workloads provide \sysname{} with greater capacity to overlap model loading and inference.

\begin{figure}
    \centering
    \includegraphics[width=.89\linewidth]{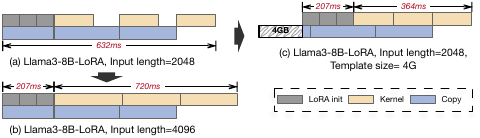}
    \vspace{-2mm}
    \caption{Breakdown of \sysname{} under different conditions.}
    \vspace{-2mm}
    \label{fig:llama8b-lora-breakdown}
\end{figure}
\paragraph{Improvement breakdown in \sysname{}.}
\autoref{fig:llama8b-lora-breakdown} summarizes \sysname{}’s improvement breakdown using the example of Llama3-8B with LoRA enabled.
The optimizations include three key steps: kernel code segments are proactively loaded before invocation, only the LoRA adapters are initialized during invocation, and model loading is overlapped with both adapter initialization and model inference.
Three distinct cases are observed: when the input sequence length is 2k, the TTFT is 632ms, dominated by model loading;
increasing the template size to 4GB reduces the TTFT to 571ms, dominated by model inference;
and increasing the input sequence length to 4096 raises the TTFT to 927ms, also dominated by model inference.
These results highlight how workload (input length) and template size influence performance, with TTFT bottlenecks shifting between model loading and inference.

\paragraph{TTFT with distributed inference.}

\begin{figure}
    \centering
    \includegraphics[width=1.0\linewidth]{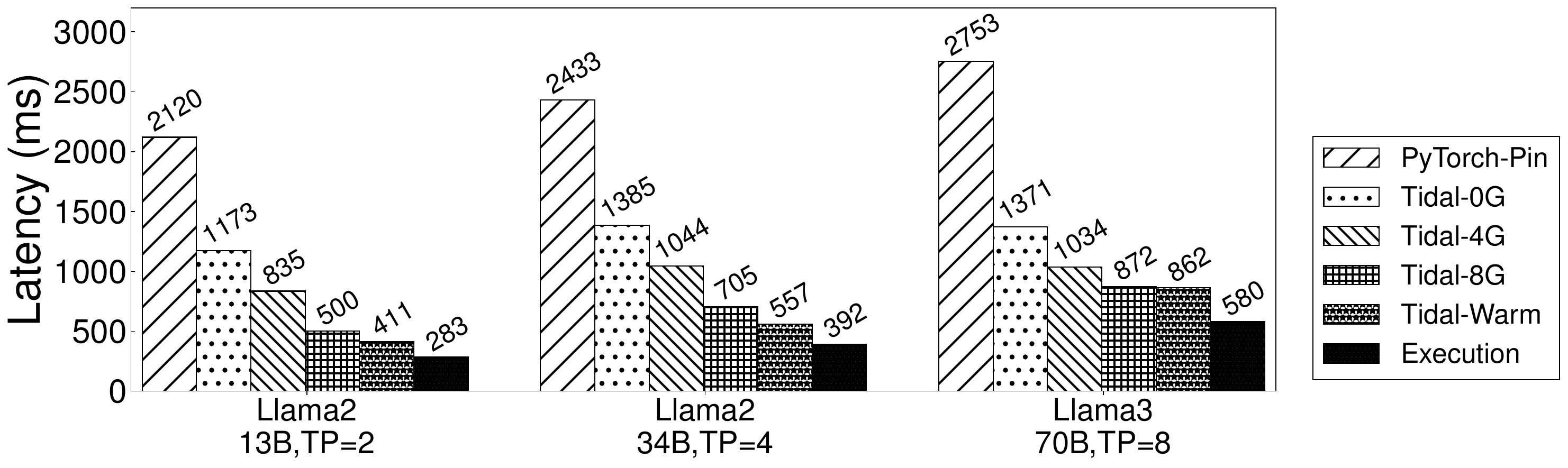}
    \vspace{-5mm}
    \caption{TTFT of three distributed functions using Llama-family models, with input length $=4096$ and batch size $=1$.}
    \label{fig:ttft-tp}
\end{figure}

We also evaluate the performance of \sysname{} in a distributed inference setting. Functions are defined using Llama-family models with 13 billion, 34 billion, and 70 billion parameters, running on 2, 4, and 8 A100 GPUs, respectively, on our second testbed.
Each LLM is parallelized using tensor parallelism~\cite{shoeybiMegatronLMTraining}.
We compare \sysname{} against \texttt{PyTorch-pin} and \texttt{Execution}, while \texttt{ServerlessLLM} is excluded as it does not natively support tensor parallelism.
\autoref{fig:ttft-tp} shows the TTFT of three distributed LLM functions. Compared to \texttt{PyTorch-pin}, \texttt{Tidal-0G}, \texttt{Tidal-4G}, \texttt{Tidal-8G}, and \texttt{Tidal-Warm} achieve TTFT speedups of $1.76\times\text{\textasciitilde}2.01\times$, $2.33\times\text{\textasciitilde}2.66\times$, $3.15\times\text{\textasciitilde}4.24\times$,and $3.19\times\text{\textasciitilde}5.16\times$, respectively. This consistent performance demonstrates \sysname{}’s scalability in a distributed inference environment.

\subsection{Evaluation with Real-world Traces}
\begin{table}
\scriptsize
    \centering
    \caption{LLM tasks with their average input length provided, as it significantly impacts the TTFT of each LLM invocation.}
        \label{tab:llm-tasks}
    \vspace{-2mm}
    \begin{tabular}{c||c|c|c|c}
    \hline
        \textbf{Tasks} & Mail~\cite{see-etal-2017-get}  & Conv~\cite{patelSplitwiseEfficient} & Code~\cite{patelSplitwiseEfficient} & LongBench~\cite{bai2023longbench} \\
        \hline
        \textbf{Input len.} (Avg.) & 867 & 1154 & 2048 & 6101 \\
\hline
    \end{tabular}
    \vspace{-2mm}
\end{table}
We further evaluate \sysname{} using real-world workloads on 4 servers of first testbed. Specifically, we generate 16 LLM function traces by combining real-world serverless workloads~\cite{shahradServerlessWilda} with LLM tasks~\cite{see-etal-2017-get,patelSplitwiseEfficient,bai2023longbench}.
Out of the 16 function traces, we include four replications each of Llama3-8B, Llama3-8B-LoRA, Llama2-13B, and Llama2-13B-LoRA. Each replication is associated with a specific task listed in \autoref{tab:llm-tasks}.
The function traces represent low, medium, and high invocation rates, covering a range of input sequence lengths.
Across all traces, the batch size is fixed at 1.
To ensure feasibility, the traces are scaled and accelerated to complete within 6 hours, as the original traces, sourced from a large cluster, span a duration of 7 days.
For evaluation purposes, the request timeout during scheduling is set to 60 seconds.
As detailed in \S\ref{sec:cold-start}, all models are pre-loaded into a host-side pinned-memory pool.

\begin{figure}
    \centering
    \includegraphics[width=\linewidth]{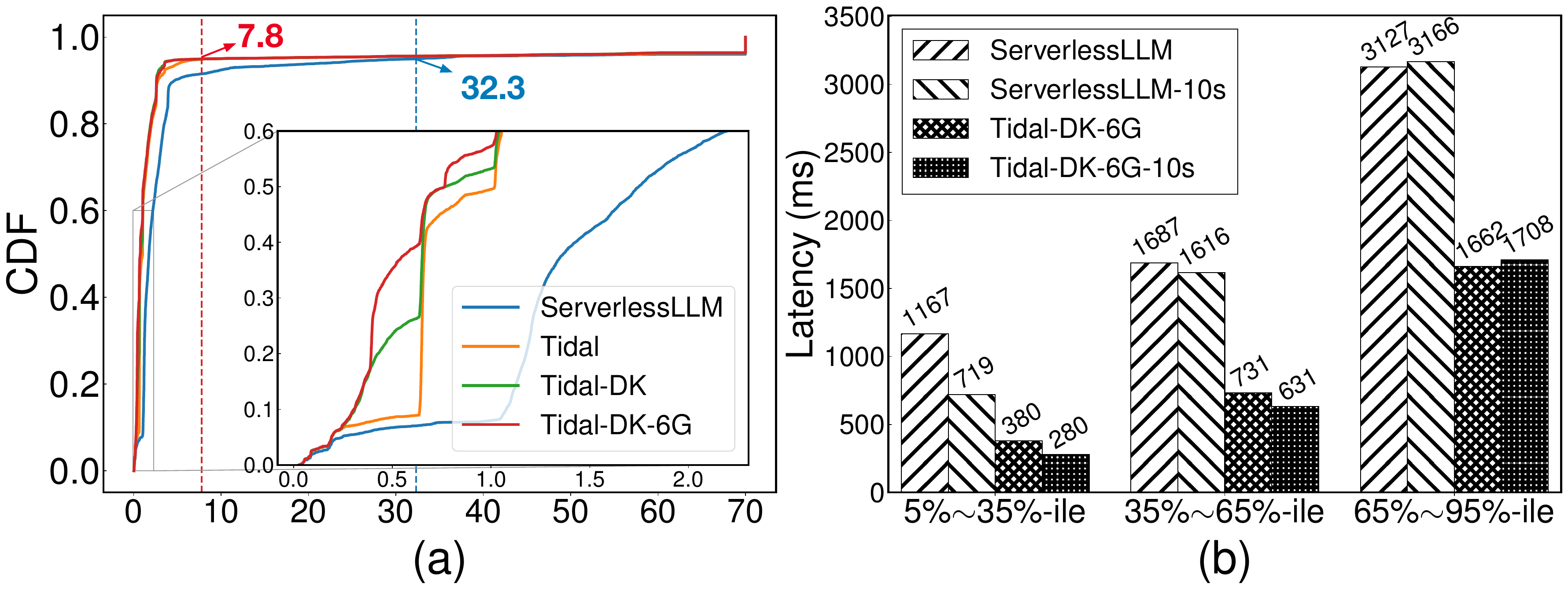}
    \vspace{-6mm}
    \caption{(a) CDF of under real-world workloads (b) Average latency of different percentile stages.}
    \vspace{-1mm}
    \label{fig:cluster}
\end{figure}

We compare \sysname{} with ServerlessLLM~\cite{fuServerlessLLMLowlatency}, the only publicly available end-to-end FaaS solution.
We first set the keep-alive interval to the model loading time following ServerlessLLM.
\autoref{fig:cluster}-(a) presents the results for ServerlessLLM alongside three variants of \sysname{}:
\texttt{Tidal} sets the template size of all functions to zero.
\texttt{Tidal-DK} extends \texttt{Tidal} by enabling keep-alive for dynamic functions.
\texttt{Tidal-DK-6G} further enhances \texttt{Tidal-DK} by selecting function 4 traces and increasing their template size to 6GB of memory on 2GPUs.
Since 8 GPUs are used for evaluation, 6GB of each selected GPU’s memory is allocated for caching the function templates.
The adjustments of template size are guided by \autoref{eq:template-size}.

Compared to \texttt{ServerlessLLM}, \texttt{Tidal} reduces the $95\%$-ile latency of TTFT by $76.0\%$, as \sysname{} significantly reduces GPU-side cold start across all LLM functions.
As illustrated in the amplified CDF, the variants of \sysname{} progressively reduce function latency, with each variant outperforming the previous one.
With keep-alive enabled for dynamic LLM functions, \texttt{Tidal-DK} effectively avoids the cold starts of dynamic LLM invocations with high invocation rates.
In \texttt{Tidal-DK-6G}, 6GB of GPU memory is allocated for storing templates.
However, this memory usage does not impact the invocation of other functions, indicating a high template density.
Additionally, by increasing the template size for four functions, \texttt{Tidal-DK-6G} further reduces their cold start latency.
Overall, \texttt{Tidal-DK-6G} achieves the best performance.

To further evaluate \sysname{}’s cluster-wide performance, we increased the keep-alive interval to 10 seconds.
\autoref{fig:cluster}-(b) shows the average latency across various percentile stages.
Under different keep-alive configurations, \sysname{} consistently outperforms ServerlessLLM at all percentile stages, demonstrating its robustness as a FaaS framework for LLMs.

\subsection{Optimizations}
\label{sec:eval:optimization}
\begin{figure}
    \centering
    \includegraphics[width=\linewidth]{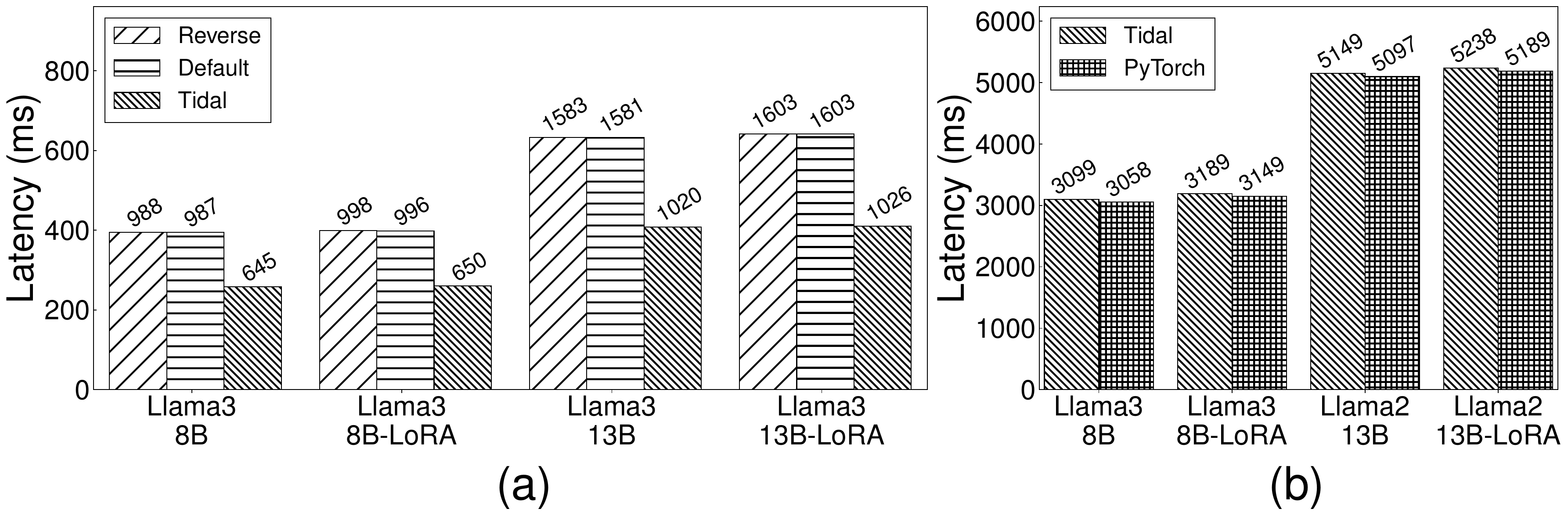}
    \vspace{-6mm}
    \caption{(a) TTFT of LLM invocations with different weight loading orders; (b) Decoding latency of LLM invocations with 99 output tokens.}
    \vspace{-4mm}
    \label{fig:eval:optimization}
\end{figure}
\paragraph{Loading order of model weights.}
We evaluate \sysname{} under different weight-loading orders, with the results shown in \autoref{fig:eval:optimization}-a.
\texttt{Tidal} denotes the access order traced by \sysname{}.
\texttt{Reverse} represents the reverse of \texttt{Tidal}, while \texttt{Default} corresponds to the initialization order of the weights.
Compared to \texttt{Reverse} and \texttt{Default}, \texttt{Tidal} achieves performance improvements of $1.55\times$ and $1.54\times$, respectively.
The similar performance of \texttt{Default} and \texttt{Reverse} attributes to the fact that many LLMs, such as Llama2~\cite{touvronLlama2}, share the weights of their embedding layer (the first layer) with the final output layer.
Moreover, this weight tensor is initialized and loaded by the last layer but accessed first during inference.
In contrast, \sysname{} automatically extracts such fine-grained execution paths, thus maximizing overlapping efficiency.

\paragraph{Overhead reduction in \sysname{}.}
The overhead in \sysname{} arises from two aspects.
First, proactive code segment loading increases the memory consumption of each pre-warmed process for LLM functions.
Experimental results indicate that \sysname{} raises memory usage per pre-warmed process from 270MB to 350MB and extends process pre-warming time from 830ms to 1070ms.
Such overhead is acceptable relative to the substantial requirements of an LLM function invocation.
Second, \sysname{}’s runtime tracing could reduce inference performance.
In \autoref{fig:eval:optimization}-(b), we compare the LLM decoding latency of \sysname{} with native PyTorch to evaluate this impact. During this phase, the model weights are already loaded into memory, but \sysname{}’s runtime tracing remains active to ensure correctness guarantees.
Compared to native PyTorch, \sysname{} incurs an overhead of less than $1.2\%$.

\paragraph{Memory pool in template server.}
\begin{table}
\footnotesize
    \centering
    \caption{TTFT (ms) of Llama2-70B on 8 A100 GPUs, with and without weight tensor merging.}
    \label{tab:memory-merge}
    \begin{tabular}{c||c|c|c|c|c|c}
    \hline
      \textbf{Input len.}  & 512 & 1024 & 2048 & 4096 & 8192 & 16384\\
        \hline
      \textbf{No Merge}  & 1366 & 1370 & 1371 & 1541 & 2203 & 4084\\
         \hline
        \textbf{Merge} & 1363 & 1364 & 1366 & 1381 & 1539 & 3406\\
         \hline
    \end{tabular}
\end{table}
When there are too many weight tensors, \sysname{}’s template server merges them into fewer weight tensors to improve overlapping efficiency. For instance, while Llama2-70B initializes 1,200 weight tensors, \sysname{} merges them into just 300 tensors.
\autoref{tab:memory-merge} compares the TTFT of Llama2-70B with and without weight tensor merging.
Without tensor merging, the overhead increases with input length but stabilizes at 600 milliseconds.
By leveraging tensor merging, \sysname{} achieves strong scalability, even as LLMs scale to extremely large sizes.

\subsection{Security Analysis in \sysname{}}
\sysname{} shares static initialization across different invocations, ensuring that this initialization does not include any request-specific content. Furthermore, \sysname{}’s runtime tracer guarantees that all weight tensors are shared in a copy-on-write manner during inference.
A potential data leak could occur if application developers’ customized GPU kernels bypass \sysname{}’s runtime tracer.
To address this, \sysname{} requires developers to register their customized GPU kernels in PyTorch for tracing.
A static code analyzer could be employed to identify and mitigate such issues.

\section{Related Work}
\paragraph{Optimizations of Cold Start.}
Several studies~\cite{liTetrisMemoryefficienta,hongOptimusWarming,suiPrewarmingNot} have explored optimizing cold starts in FaaS frameworks for deep learning inference.
Tetris~\cite{liTetrisMemoryefficienta} reduces memory footprints through tensor sharing to warm more functions, while Optimus~\cite{hongOptimusWarming} reuses model structures across functions to optimize function loading.
However, thet primarily target deep learning models~\cite{simonyan2014very,heDeepResidual}, and focus exclusively on CPU-based inferences.
In contrast, \sysname{} addresses GPU-side cold starts for LLM functions, a critical and overlooked issue.

InstaInfer~\cite{suiPrewarmingNot} addresses GPU-side challenges by employing preloading techniques but falls short of fully eliminating cold starts due to its load-dependent design.
ServerlessLLM~\cite{fuServerlessLLMLowlatency} optimizes data transfers from storage to host memory and employs a locality-driven scheduler for reducing cold start for LLM functions.
However, ServerlessLLM neglects GPU-side cold starts, missing opportunities to overlap host-to-GPU data transfers with model inference, proactively load critical code segments, and tackle the cold starts of dynamic LLM functions.

\paragraph{Template-start in CPU-only FaaS}
Template-start techniques~\cite{duCatalyzerSubmillisecond,liRunDLightweight} have been widely adopted in CPU-only FaaS frameworks to eliminate cold starts for new invocations.
Among these, Catalyzer~\cite{duCatalyzerSubmillisecond} was the first to propose launching new invocations from an existing template, effectively bypassing function initialization.
Rund~\cite{liRunDLightweight} further improves template deployment density by introducing lightweight secure container runtime.
Inspired by these works, \sysname{} goes further by generating adaptive function templates based on traced fine-grained execution paths, rather than simply reusing an existing template.

\paragraph{Serverless Scheduling for Deep Learning.}
Substantial works~\cite{ali2020batch,ali2022optimizing,yangINFlessNative,zhangMArkExploiting,romero2021infaas} have explored integrating deep learning applications into serverless environments with advanced scheduling techniques.
Some efforts~\cite{ali2020batch,zhangMArkExploiting,yang2022infless,ali2022optimizing} enable dynamic batching for deep learning inferences within serverless frameworks.
INFaaS~\cite{romero2021infaas} further introduces the concept of model-less architectures for automated management of model variants.
These advanced scheduling techniques are complementary to \sysname{} and could be integrated to further enhance the performance of LLM functions.

\section{Conclusion}
In this paper, we presented \sysname{}, an optimized Function-as-a-Service (FaaS) framework designed to address the challenges of serving Large Language Model (LLM) applications within FaaS environments.
By tracing fine-grained execution paths to generate adaptive function templates, \sysname{} effectively overcomes GPU-side cold start issues that hinder existing frameworks.
Our extensive evaluations highlight \sysname{}’s ability to significantly reduce cold start latency by $1.79\text{\textasciitilde}2.11\times$ and improve the $95\%$-ile TTFT by $76.0\%$, compared to state-of-the-art solutions.
These results confirm \sysname{}’s potential as a robust and efficient FaaS framework for the growing demands of LLM workloads.

\bibliographystyle{unsrt}
\bibliography{references/others,references/papers}

\end{document}